\documentclass[aps,pra,twocolumn,showpacs,floatfix]{revtex4}
\usepackage{epsfig}
\usepackage{graphicx}
\usepackage{dcolumn}
\usepackage{longtable}
\usepackage{amsthm,amsmath}

\begin{document}

\preprint{\today}

\title{Significance of distinct electron correlation effects in determining the P,T-odd electric dipole moment of $^{171}$Yb}

\author{B. K. Sahoo \footnote{Email: bijaya@prl.res.in}}
\affiliation{Atomic, Molecular and Optical Physics Division, Physical Research Laboratory, Navrangpura, Ahmedabad 380009, India}
\author{Yashpal Singh }
\affiliation{Graduate School of EEWS, Korea Advanced Institute of Science and Technology (KAIST), 291 Daehak-ro 34141, Daejeon, South Korea}

\begin{abstract}

Parity and time-reversal violating electric dipole moment (EDM) of $^{171}$Yb is calculated accounting for the electron correlation 
effects over the Dirac-Hartree-Fock (DHF) method in the relativistic Rayleigh-Schr\"odinger many-body perturbation theory, with the 
second (MBPT(2) method) and third order (MBPT(3) method) approximations, and two variants of all-order relativistic many-body approaches,
in the random phase approximation (RPA) and coupled-cluster (CC) method with singles and doubles (CCSD method) framework. We 
consider electron-nucleus tensor-pseudotensor (T-PT) and nuclear Schiff moment (NSM) interactions as the predominant sources that
induce EDM in a diamagnetic atomic system. Our results from the CCSD method to EDM ($d_a$) of $^{171}$Yb due to the T-PT and NSM 
interactions are found to be $d_a = 4.85(6) \times 10^{-20} \langle \sigma \rangle C_T \ |e| \ cm$  and $d_a=2.89(4) \times 
10^{-17} {S/(|e|\ fm^3)}$, respectively, where $C_T$ is the T-PT coupling constant and $S$ is the NSM. These values differ significantly from 
the earlier calculations. The reason for the same has been attributed to large correlation effects arising through non-RPA type of interactions
among the electrons in this atom that are observed by analyzing the differences in the RPA and CCSD results. This has been further scrutinized 
from the MBPT(2) and MBPT(3) results and their roles have been demonstrated explicitly.
\end{abstract}


\maketitle

\section{Introduction}

Possible existence of intrinsic electric dipole moments (EDMs) of non-degenerate quantum systems like atoms and molecules can signify
for the violations of both parity (P) and time-reversal (T) symmetries (P,T-odd) \citep{khriplovich, pospelov, dzuba, yamanaka}. In the atomic 
sector, measurements have been performed on the $^{133}$Cs, $^{205}$Tl, $^{129}$Xe, $^{199}$Hg and $^{225}$Ra atoms which only give upper 
bounds to EDMs  \cite{murthy,chin,regan,rosenberry,griffith,graner,parker}.  Owing to the open-shell structure of  $^{133}$Cs and
$^{205}$Tl atoms, they are suitable to probe electron EDM ($d_e$) and electron-nucleus (e-N) P,T-odd pseudoscalar-scalar (PS-S) interactions. 
However, in recent past experiments on polar molecules with strong internal electric field have provided tremendous improvement on the limits on 
$d_e$ and e-N coupling-coefficient due to PS-S interactions over the atomic experiments \citep{hudson,baron}. On the other hand diamagnetic 
(closed-shell) atoms are better suitable to infer the nuclear Schiff moment (NSM) and the coupling coefficients associated with the e-N 
tensor-pseudotensor (T-PT) and scalar-pseudoscalar (S-PS) interactions. The NSM originates primarily due to the distorted charge distribution
inside the atomic nucleus caused by the P,T-odd interactions among the nucleons or from the EDMs and chromo-EDMs of the up ($\tilde{d}_u$) 
and down ($\tilde{d}_d$) quarks \cite{khriplovich,yamanaka}. At the tree level, magnitudes of these P,T-odd interactions are predicted to be 
tiny in the well celebrated standard model (SM) of particle physics. However, such P,T-odd effects are enlarged manifold in various extensions 
of SM such as multi-Higgs, supersymmetry, left-right symmetric models that are trying to address some of the today's very fundamental issues  
like observation of finiteness of neutrino masses, reasons for observing the matter-antimatter asymmetry in the Universe, existence of dark 
matter etc. \cite{dine,canetti}. Thus, the improved limits on EDMs inferred from the atomic experiments combined with accurate calculations 
can be very useful to support validity of these proposed models.

Successively, a variety of progressive experimental techniques have been used to improve the precision of EDM measurements in closed-shell 
atoms. For example, the use of spin-exchange pumped masers and a $^3$He co-magnetometer by Rosenberry and Chupp, which yields an upper limit
to Xe EDM as $d_a( ^{129}\rm{Xe})=0.7\pm 3.3(\rm{stat})\pm 0.1(\rm{sys})\times 10^{-27}$ e-cm  \cite{rosenberry}. Currently, new proposals 
to measure EDM of $^{129}$Xe are being made to take advantage of its larger spin relaxation time \cite{inoue-xe,fierlinger,schmidt}. The 
proposal by Inoue {\it et al.} \cite{inoue-xe} argues utilization of the nuclear spin oscillator technique \cite{yoshimi} to carry out 
measurement of Larmor precession with several orders lower than the available results. In the atoms like $^{223}$Rn, large enhancement of the 
EDM signal is expected owing to its deformed nucleus \cite{auerbach}. Based on this argument, an experiment to measure EDM of $^{223}$Rn has been
under progress \cite{rand,tardiff}. So far the most precise atomic EDM measurement has been performed on the $^{199}$ Hg atom, gradually improving 
its limit in two successive experiments \cite{griffith,graner}, among which the best limit has recently been reported by Graner {\it et al.}
\cite{graner}. In the earlier experiment, Griffith {\it et al.} had used a stack of four cells in such a way that electric fields were being
created in opposite directions among two middle cells and zero electric field in the outer two cells. Thus, the signal due to EDM was observed 
as a difference of the Larmor spin precession frequencies originating from the middle two cells, and combinations of these four cells were
used to measure the magnetic field. In this approach EDM of the $^{199}$Hg atom was observed as $d_a( ^{199}\rm{Hg}) = (0.49 \pm 1.29 
(\rm{stat}) \pm 0.76 (\rm{sys})) \times 10^{−29}$ e-cm \cite{griffith}. However, in the recent measurement by Graner {\it et al.} fused 
silica vapor cells containing $^{199}$Hg atoms were arranged in a stack with common magnetic field. Optical pumping was being used to 
spin-polarize the $^{199}$Hg atoms which was orthogonal to the applied magnetic field, and the Faraday rotation of near-resonant light was 
observed to determine an electric-field-induced perturbation to the Larmor precession frequency. The improved EDM value inferred from the 
above precession frequencies as $d_a( ^{199}\rm{Hg})=(−2.20 \pm 2.75(\rm{stat}) \pm 1.48 (\rm{sys})) \times 10^{−30}$ e-cm that translates to
an upper limit of $|d_a( ^{199}\rm{Hg})| < 7.4 \times 10^{−30}$ e-cm  with 95\% confidence limit, which corresponds to an improvement of 
at least an order of magnitude over the previous measurement \cite{graner}. In a breakthrough, Parker and co-workers have reported measurement 
of EDM of the radioactive element $^{225}$Ra atom for the first time \cite{parker}. Similar to $^{223}$Rn, EDM signal of $^{225}$Ra is also 
enhanced extraordinarily high due to the octupole deformation in its nucleus \cite{auerbach}. Owing to this fact, even if one could measure 
EDM of the $^{225}$Ra atom to a couple of orders larger than the $^{199}$Hg EDM, it is still advantageous to use this result to extract
out the required information more reliably. To measure EDM of the $^{225}$Ra atom, a cold-atom technique was developed to detect the spin 
precession holding the atoms in an optical dipole trap. An upper limit as $|d_a( ^{225}\rm{Ra})|< 5.0 \times 10^{−22}$ e-cm with a 95\% 
confidence level was inferred from this measurement. 

A number of calculations employing variants of relativistic atomic many-body methods have been carried out in the $^{129}$Xe, $^{223}$Rn,
$^{199}$Hg and $^{225}$Ra atoms to evaluate quantities that in combinations with the measurements can give limits on various quantities of 
fundamental interest (for more details see a recent review \cite{yamanaka}). On comparing EDM results from the latest calculations by the 
relativistic coupled-cluster (RCC) method with the previously reported values from other less sophisticated approaches, it was observed that results were almost in agreement
with each other in the $^{129}$Xe \cite{yashpal1} and $^{223}$Rn \cite{yashpal2} atoms. This suggested to us that there are strong cancellations 
among electron correlation effects in these atoms from the higher order effects. However, we had found very large differences in the results from the RCC 
method with the earlier reported calculations in the $^{199}$Hg \cite{yashpal3,sahoo1} and $^{225}$Ra atoms \cite{yashpal4}. Though detailed 
analysis on the reasons for large discrepancies in all those calculations were not given before, but we had mentioned briefly how the electron
correlation effects that do not appear through the random phase approximation (RPA) are solely responsible for bringing down the results in the latter mentioned two atoms. The other diamagnetic atom 
that is of current interest to measure EDM is the $^{171}$Yb atom \cite{liu-china}. In this proposal, it is suggested to use $^{171}$Yb 
as a co-magnetometer and a proxy for measuring EDM of the $^{225}$Ra atom. Earlier, feasibility of measuring EDM in this atom was being 
discussed extensively in Refs. \cite{takahashi,romalis,natarajan,takano} following which a number of theoretical calculations have also already 
been performed \cite{angom,dzuba2,dzuba7,dzuba9,latha}. In view of the above mentioned substantial discrepancies among the results between 
different theoretical studies in some of the atoms, it would be of vested interest to perform RCC calculations in the $^{171}$Yb atom and 
compare the obtained results with the previously reported values. Providing reliable calculations for this atom can be very useful to infer 
limits on various fundamental parameters by combining those values with the measured EDM of the $^{171}$Yb atom from the ongoing experiment 
when it comes to fruition. 

The rest of the paper is organized as follows: In the next section, we briefly mention about the theory of atomic EDMs and present  the T-PT 
and NSM interaction Hamiltonians used for the EDM calculations. Then in the next section, we describe our many-body methods and procedures 
for obtaining atomic wave functions at various levels of approximations. This is followed by discussions on our results and comparison of 
these values with the previously performed calculations. Unless stated otherwise, we use atomic units (a.u.) throughout this paper.

\section{Theory}

The P,T-odd Lagrangian for a pair of electron and nucleon (e-n) is given by \cite{pospelov}
\begin{eqnarray}
\mathcal{L}_{e-n}^{PT} &=& C_T^{e-n} \varepsilon_{\mu \nu \alpha \beta} \bar{\psi}_e \sigma^{\mu \nu} \psi_e  
\bar{\psi}_n \sigma^{\alpha \beta} \psi_n \nonumber \\ && + C_P^{e-n}  \bar{\psi}_e  \psi_e \ \bar{\psi}_n i \gamma_5 \psi_n,
\end{eqnarray}
where $\varepsilon_{\mu \nu \alpha \beta}$ is the Levi-Civita symbol and $\sigma_{\mu \nu} = \frac{i}{2}[\gamma_\mu, \gamma_\nu]$ 
with $\gamma$ being the Dirac matrices. The constants $C_T^{e-n}$ and $C_P^{e-n}$ represent couplings associated with the respective  
T-PT and S-PS e-n interactions. Here, $\psi_n$ and $\psi_e$ are the Dirac wave functions of a nucleon and an electron, respectively. In the 
non-relativistic limit, the e-n T-PT interaction Hamiltonian from the above Lagrangian yields \cite{barr1, martensson}
\begin{eqnarray}
H^{e-n}_{T-PT} &=& \frac{G_F}{\sqrt{2}} C_T^{e-n} \bar{\psi}_e \gamma_5 \sigma_{\mu \nu} \psi_e \ 
\bar{\psi}_n \iota \gamma_5 \sigma_{\mu \nu} \psi_n, 
\end{eqnarray}
where $G_F$ reads as the Fermi constant. In the atomic scale, the above equation can be further simplified to get the corresponding 
e-N T-PT interaction Hamiltonian  as
 \begin{eqnarray}
 H_{EDM}^{T-PT}=i \sqrt{2} G_FC_T \sum_e \mbox{\boldmath $\sigma_N \cdot \gamma_e$} \rho_N(r_e), 
 \end{eqnarray}
with $C_T$ being the e-N T-PT coupling constant, {\boldmath$\sigma_N$}$=\langle \sigma_N \rangle \frac{{\bf I}}{I}$ is the Pauli spinor 
of the nucleus for the nuclear spin $I$, $\rho_N(r)$ is the nuclear density and the subscripts $N$ and $e$ represent for the respective 
nucleon and electronic coordinates.

Similarly, the Lagrangian for the P,T-odd pion-nucleon-nucleon ($\pi$-n-n) interactions that contribute significantly to the 
EDMs of the diamagnetic atoms is given by \cite{pospelov}
\begin{eqnarray}
\mathcal{L}^{\pi n n}_{e-n} &=& \bar{g}_0 \bar{\psi}_n \tau^i \psi_n \pi^i + \bar{g}_1 
\bar{\psi}_n \psi_n \pi^0 \nonumber \\ && + \bar{g}_2 \big ( \bar{\psi}_n \tau^i \psi_n \pi^i - 
3 \bar{\psi}_n \tau^3 \psi_n \pi^0 \big ),
\end{eqnarray}
where the couplings $\bar{g}_i$ with the superscript $i=$ 1, 2, 3 represent for the isospin components. The corresponding e-N interaction 
Hamiltonian is given by \cite{flambaum,dzuba}
 \begin{eqnarray}
  H_{EDM}^{NSM}= \frac{3{\bf S.r}}{B_4} \rho_N(r),
 \end{eqnarray}
where ${\bf S}=S \frac{{\bf I}}{I}$ is the NSM and $B_4=\int_0^{\infty} dr r^4 \rho_N(r)$. The magnitude of NSM $S$ is given by 
\cite{haxton, ban, jesus}
\begin{eqnarray}
S = g_{\pi n n} \times (a_0 \bar{g}_{\pi n n}^{(0)} + a_1 \bar{g}_{\pi n n}^{(1)} + a_2 \bar{g}_{\pi n n}^{(2)}),
\end{eqnarray}
where $g_{\pi nn} \simeq 13.5$ is the CP-even $\pi$-n-n coupling constant, $a_i$ are the polarization parameters of the nuclear charge 
distribution that can be computed to a reasonable accuracy using the Skyrme effective interactions in the Hartree-Fock-Bogoliubov 
mean-field method \cite{engel}, and $\bar{g}_{\pi n n}^{(i)}$s with $i=$ 1, 2, 3 represent for the isospin components of the CP-odd 
$\pi$-n-n coupling constants. These couplings are related to the chromo-EDMs of up quark ($\tilde{d}_u$) and down quark ($\tilde{d}_d$) 
as $\bar{g}_{\pi n n}^{(1)} \approx 2 \times  10^{-12} \times (\tilde{d}_u - \tilde{d}_d)$ \cite{pospelov,pospelov1} and 
$\bar{g}_{\pi n n}^{(0)}/ \bar{g}_{\pi n n}^{(1)} \approx 0.2 \times (\tilde{d}_u + \tilde{d}_d)/(\tilde{d}_u -\tilde{d}_d)$ 
\cite{pospelov,dekens}, where $\tilde{d}_u$ and $\tilde{d}_d$ are scaled to $10^{-26}$ e-cm. Also, it yields a relation 
with the quantum chromodynamics (QCD) parameter ($\bar{\theta}$) by $|\bar{g}_{\pi n n}^{(1)}|=0.018(7) \bar{\theta}$ \cite{dekens}. From 
the nuclear calculations, one can obtain $S \simeq (1.9d_n+0.2d_p)$ fm$^2$ \cite{dmitriev}. Thus, it is necessary to obtain accurate values 
of $C_T$ and $S$ by combining atomic calculations with the experimental EDM result to infer magnitudes of the above fundamental
parameters reliably.

\section{Method of calculations}

The EDM of an atomic system in its ground state is given by 
\begin{eqnarray}
 d_{a} = \frac{\langle \Psi_0 | D | \Psi_0 \rangle}{\langle \Psi_0 | \Psi_0 \rangle },
 \label{edmeq}
\end{eqnarray}
where $D$ is the electric dipole (E1) operator and ($|\Psi_0 \rangle$) is the ground state wave function corresponding to the atomic
Hamiltonian containing both the electromagnetic and P,T-odd weak interactions. Since atoms are spherically symmetric, we use the spherical 
polar coordinate system  to determine atomic wave functions. In this case, operators are expressed in form of multiple expansion and parity 
is treated as a good quantum number. Thus, mixture of parities in the wave functions due to both the electromagnetic and weak interactions 
are done explicitly when required. For this reason, we evaluate the atomic wave functions first by considering only the electromagnetic 
interactions where parities of the atomic orbitals are still preserved. Then these wave functions are perturbed to the first order due to 
the P,T-odd operators because of which parity mixing among the atomic orbitals are carried out explicitly. This is done by expressing 
the atomic Hamiltonian as
\begin{eqnarray}
 H = H^{at} + \lambda H^{PT} ,
\end{eqnarray}
 where $H^{at}$ represents the Hamiltonian part that accounts only the electromagnetic interactions and $\lambda H^{PT}$ corresponds to 
one of the considered P,T-odd Hamiltonians with $\lambda$ representing either $S$ or $C^T$ depending upon the undertaken P,T-odd Hamiltonian. 
In this framework, atomic wave function $|\Psi_0 \rangle$ can be expressed as
\begin{eqnarray}
 |\Psi_0 \rangle \approx |\Psi_0^{(0)} \rangle + \lambda |\Psi_0^{(1)} \rangle  ,
\end{eqnarray}
where $| \Psi_0^{(0)} \rangle$ and $|\Psi_0^{(1)} \rangle$ are the wave functions due to $H^{at}$ and its first order correction due 
to $\lambda H^{PT}$, respectively.  Following this Eq. (\ref{edmeq}) can be approximated to
\begin{eqnarray}
 d_{a} & \simeq & 2 \lambda \frac{\langle \Psi_0^{(0)}|D|\Psi_0^{(1)} \rangle}{\langle \Psi_0^{(0)}|\Psi_0^{(0)} \rangle}.
 \label{eqed}
\end{eqnarray}
To infer $S$ and $C_T$ values from the measured $d_{a}$ result, it is imperative to determine
\begin{eqnarray}
 {\cal R} = d_{a} / {\lambda} &=& 2 \frac{\langle \Psi_0^{(0)}|D|\Psi_0^{(1)} \rangle}{\langle \Psi_0^{(0)}|\Psi_0^{(0)} \rangle} .
\label{eqptt}
 \end{eqnarray}
The first order perturbed wave function $|\Psi^{(1)} \rangle$ can be obtained as a solution of the following inhomogeneous equation
\begin{eqnarray}
(H^{at}-E_0^{(0)}) |\Psi_0^{(1)} \rangle &=& (E_0^{(1)}- H^{PT}) |\Psi_0^{(0)} \rangle \nonumber \\
                                         &=& - H^{PT} |\Psi_0^{(0)} \rangle ,
\label{eq4}
\end{eqnarray}
where $E_0^{(0)}$ and $E_0^{(1)}$ are the zeroth and the first order perturbed energies of the ground state. Here  $E_0^{(1)}$ vanishes 
owing to odd-parity nature of $H^{PT}$. 

It is worth mentioning here that, one can obtain ground state E1 polarizability ($\alpha_d$) of the atomic system by using $\lambda |\Psi_0^{(1)} \rangle$
as the first order order perturbed wave function due to the operator $D$ in Eq. (\ref{eqed}), which can be evaluated by simply  
replacing $\lambda H^{PT}$ by operator $D$ in the above inhomogeneous equation. Conventionally, robustness of a many-body method can be judged by 
its potential to reproduce experimental results. Though a precise experimental value of $\alpha_d$ for Yb is not available, we still 
carry out calculations of $\alpha_d$ of the $^{171}$Yb atom by employing the considered many-body methods and compare our result with the previously 
available results from other theoretical studies to get some assurance on the accuracies of our calculated ${\cal R}$ values. 

In fact, calculating atomic wave functions accurately due to the electromagnetic interactions by allowing only one photon exchange,
even in the non-covariant form approximation, is also strenuous owing to the two-body form of the electron-electron interaction potential. We 
consider the Dirac-Coulomb (DC) Hamiltonian as $H^{at}$, which is given by
\begin{eqnarray}
H^{at} &=& \sum_i \left [ c\mbox{\boldmath$\alpha$}_i\cdot \textbf{p}_i+(\beta_i -1)c^2 + V_n(r_i) + \sum_{j>i} \frac{1}{r_{ij}} \right ] \nonumber \\
\end{eqnarray}
with $\mbox{\boldmath$\alpha$}$ and $\beta$ are the usual Dirac matrices and $V_n(r)$ represents for the nuclear potential that is
evaluated assuming the Fermi-charge distribution, for calculating $|\Psi_0^{(0)} \rangle$. 

In this work, we consider relativistic second order many-body perturbation theory (MBPT(2)) and third order many-body perturbation theory
(MBPT(3)) in the Rayleigh-Schr\"odinger approach, RPA and RCC methods for calculating $\alpha_d$ and ${\cal R}$ values. To demonstrate relations 
among these methods, we discuss on the formulation of these methods briefly by starting with the common reference wave function 
$\vert \Phi_0 \rangle$, which is obtained here using the Dirac-Hartree-Fock (DHF) method, by expressing as
\begin{eqnarray}
 |\Psi_0^{(0)} \rangle = \Omega^{(0)} | \Phi_0 \rangle
\end{eqnarray}
and 
\begin{eqnarray}
 |\Psi_0^{(1)} \rangle = \Omega^{(1)} | \Phi_0 \rangle ,
\end{eqnarray}
where $\Omega^{(0)}$ and $\Omega^{(1)}$ are known as wave operators that account for the neglected residual electromagnetic interactions
($V_{es}$) in the DHF method and $V_{es}$ with the considered weak interactions to first order, respectively. 

In the MBPT(n) method, we expand the wave operators as
\begin{eqnarray}
|\Psi_0^{(0)} \rangle = \sum_{k}^n \Omega^{(k,0)} |\Phi_0 \rangle,
\end{eqnarray}
where $\Omega^{(k,0)}$ is the wave operator with $k$ and zero orders of $V_{es}$ and $H^{PT}$ perturbations, respectively. The first 
order correction to $ |\Psi_0^{(0)} \rangle$ due to $H^{PT}$ in the MBPT(n) method is then expressed as
\begin{eqnarray}
 |\Psi_0^{(1)} \rangle  &=& \sum_{k}^{n-1} \Omega^{(k,1)} |\Phi_0 \rangle .
\end{eqnarray}
Amplitudes corresponding to both the unperturbed and perturbed wave operators are obtained using the generalized Bloch equations as \cite{lindgren}
\begin{eqnarray}
 [\Omega^{(k,0)},H_0 ] P &=& Q V_{es} \Omega^{(k-1,0)}P  \nonumber \\ && -
 \sum_{m=1 }^{k-1} \Omega^{(k-m,0)} P V_{es} \Omega^{(m-1,0)}P \ \  \  \  \
\end{eqnarray}
and
\begin{eqnarray}
 [\Omega^{(k,1)},H_0 ]P &=& QV_{es} \Omega^{(k-1,1)}P + Q H^{PT} \Omega^{(k,0)}P 
\nonumber \\ && - \sum_{m=1 }^{k-1} \big ( \Omega^{(k-m,1)}
 P V_{es} \Omega^{(m-1,0)}P \nonumber \\ && - \Omega^{(k-m,1)}P H^{PT} \Omega^{(m,0)}P \big ),
\label{eq44}
\end{eqnarray}
where $P= |\Phi_0 \rangle \langle \Phi_0 |$ and $Q=1-P$. It implies that $\Omega^{(0,0)}=1$, $\Omega^{(1,0)}=\sum_I
\frac{ \langle \Phi_I | V_{es} | \Phi_0 \rangle} { E_I^{(0)} - E_0^{(0)}} = 0$ and $\Omega^{(0,1)}= \sum_I 
\frac{ \langle \Phi_I | H^{PT} | \Phi_0 \rangle} { E_I^{(0)} - E_0^{(0)}}$. Here $|\Phi_I\rangle$ with DHF energy $E_I^{(0)}$ is an 
excited state with respect to $|\Phi_0\rangle$ and $E_0^{(0)}$ is the sum of DHF single particle energies. We implement this method 
using the Goldstone diagrams adopting normal ordering of second quantization operators that define excitations and de-excitation 
processes from $|\Phi_0 \rangle$ considering it as the Fermi vacuum. Though this approach is convenient to implement, but number 
of diagrams increase rapidly from the MBPT(2) to MBPT(3) method (7 to more than 200 diagrams). Thus, it is challenging to go beyond 
the MBPT(3) method. However, behavior of various correlation effects can be investigated explicitly through these approximations. Here, we 
have applied these methods to explain the reasons for the discrepancies between the results obtained using the RPA and RCC methods.

In the RPA, the wave operators are approximated to
\begin{eqnarray}
\Omega^{(0)} & \approx &  1
\label{rpawaveop0}
\end{eqnarray} 
and 
\begin{eqnarray}
\Omega^{(1)} & \approx &  \sum_k^{\infty} \sum_{p,a} \Omega_{p,a}^{(k, 1)} \nonumber \\
    &=& \sum_{k=1}^{\infty} \sum_{pq,ab} \Big { ( } \frac{[\langle pb | \frac{1}{r_{ij}}  | aq \rangle 
- \langle pb | \frac{1}{r_{ij}} | qa \rangle] \Omega_{b, q}^{(k-1,1)} } {\epsilon_p - \epsilon_a}  \nonumber \\ 
&& + \frac{ \Omega_{b, q}^{{(k,1)}^{\dagger}}[\langle pq | \frac{1}{r_{ij}} | ab \rangle - \langle pq | \frac{1}{r_{ij}} | ba \rangle] 
}{\epsilon_p-\epsilon_a} \Big {) },
\label{rpawaveop1}
\end{eqnarray} 
where sum over $a$ and $p$ represent replacement of an occupied orbital $a$ by a virtual orbital $p$ in $|\Phi_0 \rangle$, corresponding 
to a class of single excitations. Formulation of wave operator in this approach encapsulates the core-polarization effects to all orders, 
which play dominant role in determining the investigated properties in this work.

In the RCC method, we express the unperturbed wave operator as
\begin{eqnarray}
 \Omega^{(0)} &=& e^{T^{(0)}}  
 \label{ccwaveop1}
\label{eq32}
\end{eqnarray}
and the first order perturbed wave operator as
\begin{eqnarray}
  \Omega^{(1)}   &=& e^{T^{(0)}}  T^{(1)} , 
  \label{ccwaveop2}
\label{eq33}
\end{eqnarray} 
where $T^{(0)}$ and $T^{(1)}$ are referred to as the excitation operators that produce excited state configurations after operating upon 
$|\Phi_0 \rangle$ due to $V_{es}$ and due to $V_{es}$ along with the perturbed $H^{PT}$ operator, respectively. The amplitudes of these RCC 
excitation operators are evaluated by solving the equations
\begin{eqnarray}
 \langle \Phi_0^{*}|\overline{H}_N^{at}|\Phi_0\rangle&=&0
 \label{eq36}
 \end{eqnarray}
and
\begin{eqnarray}
\langle \Phi_0^{*}|\overline{H}_N^{at}T^{(1)}|\Phi_0\rangle&=&-\langle \Phi_0^{*}|\overline{H}_N^{PT}|\Phi_0\rangle ,
  \label{eq37}
\end{eqnarray}
where the subscript $N$ represents normal ordered form of the operator, $\overline{O}=(Oe^{T^{(0)}})_{con}$ with $con$
means only the connected terms are allowed and $| \Phi_0^{*} \rangle$ corresponds to the excited configurations with respect 
to $| \Phi_0 \rangle$. In our calculations, we only consider the singly and doubly excited configurations by defining
\begin{eqnarray}
T^{(0)} &=& T_1^{(0)} + T_2^{(0)} \ \ \ \ \text{and} \ \ \ \ 
T^{(1)} = T_1^{(1)} + T_2^{(1)} ,
\label{eq35}
\end{eqnarray}
which is known as the CCSD method in the literature. The difficult part in this method is to store and compute the reduced matrix element 
of $H_N^{PT} \otimes T_2^{(0)}$, which is odd in parity, as it involves coupled tensor products in the spherical coordinate system. 

After obtaining amplitudes of the wave operators in different approaches, we obtain the DHF value of ${\cal R}$ as 
\begin{eqnarray}
 {\cal R}  &=& 2 \langle \Phi_0| {\Omega^{(0,0)}}^{\dagger} D \Omega^{(0,1)} |\Phi_0 \rangle \nonumber \\
 & =& 2 \sum_I \frac{  \langle \Phi_0| D| \Phi_I \rangle \langle \Phi_I | H^{PT} | \Phi_0 \rangle} { E_I^{(0)} - E_0^{(0)}} .
\end{eqnarray}
In the MBPT(n) method, we evaluate 
\begin{eqnarray}
{\cal R} &=& 2 \frac{\sum_{k=0}^{n-1} \langle \Phi_0| {\Omega^{(n-k,0)}}^{\dagger} D \Omega^{(k,1)} |\Phi_0 \rangle}
{ \sum_{l=0}^{n-1} \langle \Phi_0| {\Omega^{(n-l,0)}}^{\dagger} \Omega^{(l,0)} |\Phi_0 \rangle} \nonumber \\
    & \approx & 2 \sum_{k=0}^{n-1} \langle \Phi_0| ({\Omega^{(n-k,0)}}^{\dagger} D \Omega^{(k,1)})_{con} |\Phi_0 \rangle .
\end{eqnarray}
Thus, expression in MBPT(2) method is given by
\begin{eqnarray}
{\cal R} &\approx& 2 \langle \Phi_0 | [\Omega^{(0,0)}+\Omega^{(1,0)}]^{\dagger} D [\Omega^{(0,1)}+\Omega^{(1,1)}]|\Phi_0 \rangle \nonumber \\
& \approx & 2 \langle \Phi_0| D\Omega^{(0,1)} + D\Omega^{(1,1)} + {\Omega^{(1,0)}}^{\dagger} D\Omega^{(0,1)} \nonumber \\&&  +
{\Omega^{(1,0)}}^{\dagger} D\Omega^{(1,1)} |\Phi_0 \rangle  ,
\label{eq20}
\end{eqnarray}
and in the MBPT(3) method it is given by
\begin{eqnarray}
 {\cal R}  &\approx& 2 \langle \Phi_0 | [\Omega^{(0,0)}+\Omega^{(1,0)}+\Omega^{(2,0)}]^{\dagger} D \nonumber \\ && \times[\Omega^{(0,1)}+\Omega^{(1,1)}+\Omega^{(2,1)}]|\Phi_0 \rangle \nonumber \\
& \approx &  2 \langle \Phi_0| D\Omega^{(0,1)} + D\Omega^{(1,1)}+D\Omega^{(2,1)} + {\Omega^{(1,0)}}^{\dagger} D\Omega^{(0,1)}  \nonumber \\ && +
{\Omega^{(1,0)}}^{\dagger} D\Omega^{(1,1)} +{\Omega^{(2,0)}}^{\dagger} D\Omega^{(0,1)}|\Phi_0 \rangle  .
\label{eq21}
\end{eqnarray}  
This expression clearly indicates on the complexity in the calculations with the consideration of higher order terms through the MBPT(n) methods.

In the all order RPA method, we get 
\begin{eqnarray}
 {\cal R}  &=& 2 \langle \Phi_0| {\Omega^{(0,0)}}^{\dagger} D \Omega_{RPA}^{(1)} |\Phi_0 \rangle \nonumber \\
         &=& 2 \langle \Phi_0| D \Omega_{RPA}^{(1)} |\Phi_0 \rangle .
\end{eqnarray}
In fact, this method is straightforward to implement and requires much less computational time to obtain the ${\cal R}$ values than 
the MBPT(3) method. Since it is able to capture the electron core-polarization effects to all orders, one would expect to get 
reasonably accurate values using RPA than the MBPT(2) and MBPT(3) methods.

The CCSD method should give the most accurate results for ${\cal R}$ than all the employed methods as it subsumes contributions arising 
through the RPA method as well as accounts for other types of correlation effects, such as the electron pair-correlation effects, to all orders 
which are arising in the MBPT(3) method as the lowest order non-RPA type of contributions. Importantly, all these correlation effects are 
coupled through the RCC amplitude solving equations as in the real physical situation. In this approach, we evaluate as
\begin{eqnarray}
 {\cal R} &=& 2 \frac{\langle\Phi_0 | e^{T^{\dagger (0)}} D e^{T^{(0)}} T^{(1)} | \Phi_0 \rangle }
                  {\langle\Phi_0 | e^{T^{\dagger (0)}} e^{T^{(0)}} | \Phi_0 \rangle } \nonumber \\
  &=& 2 \langle\Phi_0 |(\overline{D}^{(0)} T^{(1)})_{con}|\Phi_0 \rangle,
\label{eq38}
\end{eqnarray}
where $\overline{D}^{(0)} = e^{T^{\dagger{(0)}}}De^{T^{(0)}}$ is a non-terminating series. In order to account for most of the contributions 
from $\overline{D}^{(0)}$ term, we adopt a self-consistent procedure to compute it as have been explained in our earlier works
\cite{yashpal3,sahoo1,yashpal4,yash-polz1}. 

\section{Results and Discussion}

\begin{table}[t]
\caption{Results of $\alpha_d$, ${\cal R_T}$, and ${\cal R_S}$, in units of $e a_0^3$, ($\times 10^{-20} \langle 
\sigma\rangle |e|cm$) and ($\times 10^{-17}[1/|e|fm^3]|e|cm$) respectively, for the ground state of $^{171}$Yb from the employed
many-body methods and comparison with the other studies.}
\begin{ruledtabular}
\begin{tabular}{lccc}
 Method   & $\alpha_d$&  ${\cal R_T}$& ${\cal R_S}$ \\ 
\hline 
& & \\
DHF     & 124.51 &$-$0.71  &$-$0.42   \\
MBPT(2)& 141.25  & $-2.49$ &  $-1.42$ \\    
MBPT(3)& 115.70  & $-2.34$ & $-1.34$ \\       
RPA    & 179.76  & $-3.39$ & $-1.91$  \\
CCSD & 135.50 & $-2.04$ & $-1.51$ \\
& & \\
Others  \cite{angom}       &   &   $-2.51^{\ddagger}$   &  \\
    \cite{dzuba2,dzuba7} & 179   &  $-3.4$ & $-1.91$ \\
       \cite{dzuba9} &         &         & $-2.12$ \\
       \cite{latha}  & 176.16  &         & $-1.903$ \\
       \cite{radziute} &         &         & $-2.15$ \\
       \cite{porsev} &  111.3(5)    &        & \\
       \cite{zhang}   & 143    &         & \\
       \cite{bksahoo} & 144.59(5.64) &  & \\
       \cite{dzuba}    & 141(6)  &  &  \\
       \cite{safronova} & 141(3) & & \\
Experiment \cite{miller} & 142(36) & \\
\end{tabular}
\end{ruledtabular}  
$^{\ddagger}$Sign has been changed as per the convention of this work.
\label{tab1}
\end{table}  

\begin{figure}[t]
\begin{center}
\includegraphics[width=8.5cm,height=6.0cm]{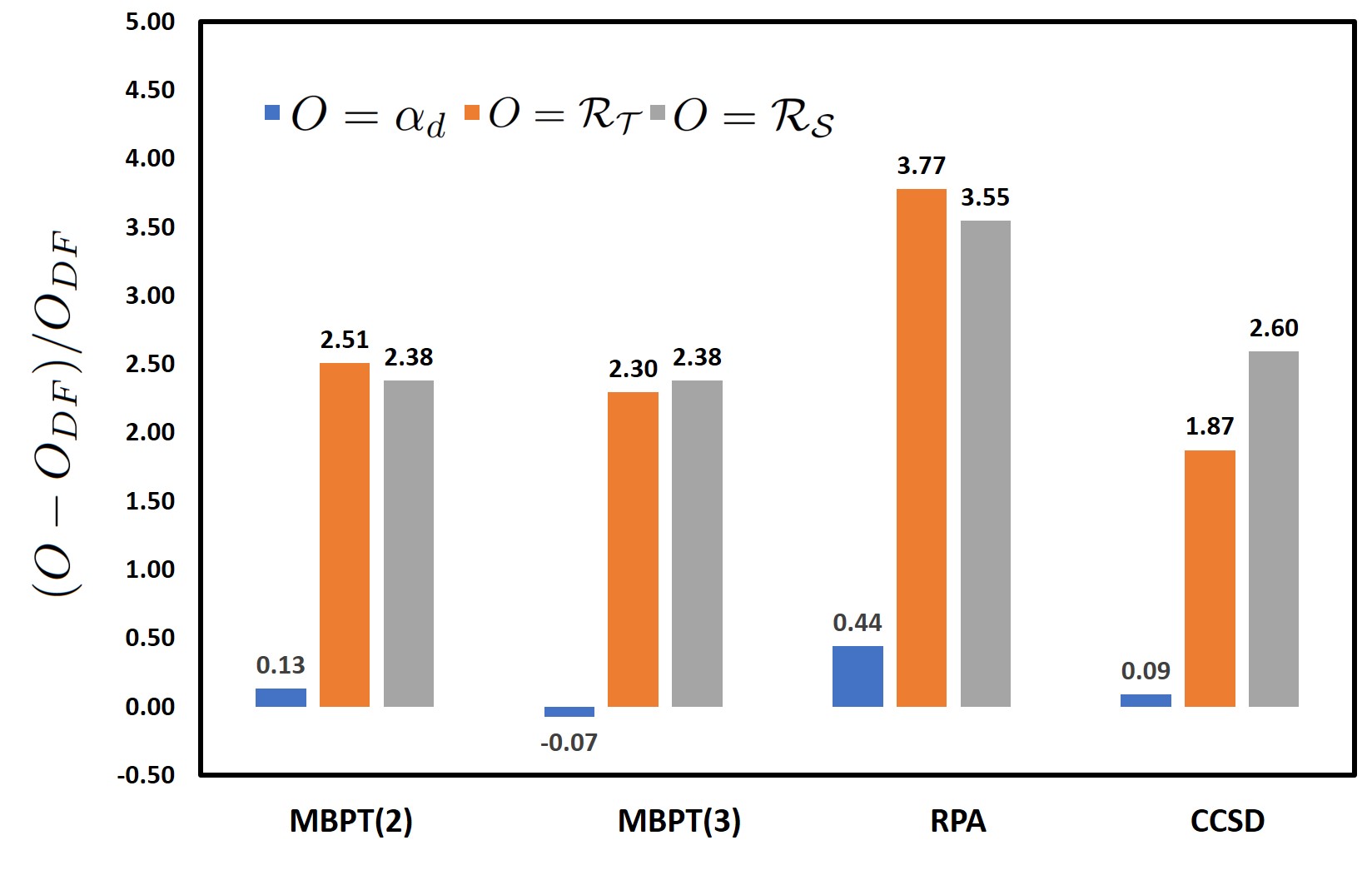}
\end{center}
\caption{Comparison of electron correlation contributions to $\alpha_d$, ${\cal R_T}$, and ${\cal R_S}$ in $^{171}$Yb at 
different levels of approximations in the many-body methods with respect to the DHF results. No scaling has been maintained in the X-axis, 
while results plotted in the Y-axis are unit-less.}
\label{fig1}
\end{figure}

\begin{figure}[t]
\begin{center}
\includegraphics[width=8.5cm,height=4.7cm]{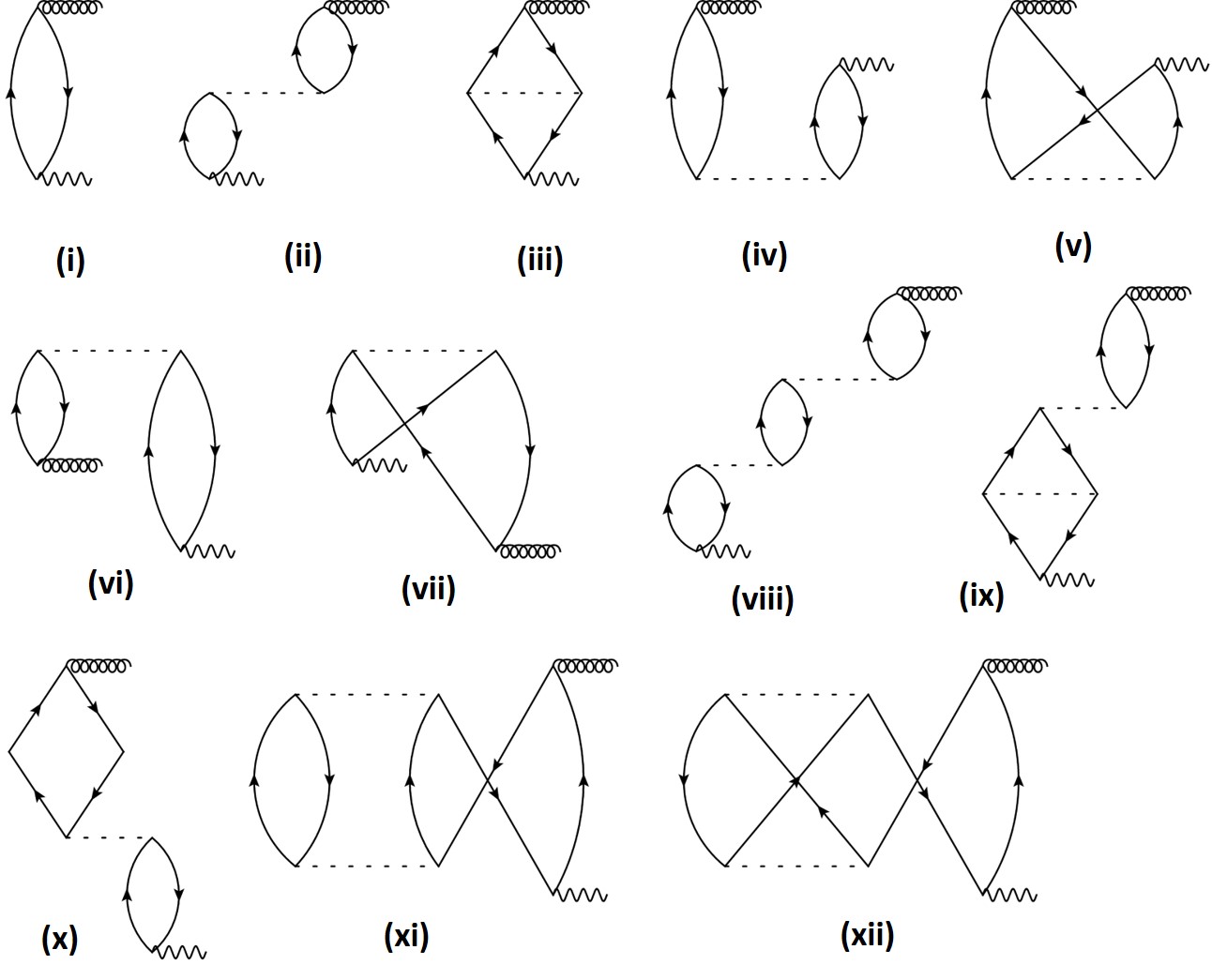}
\includegraphics[width=8.5cm,height=4.7cm]{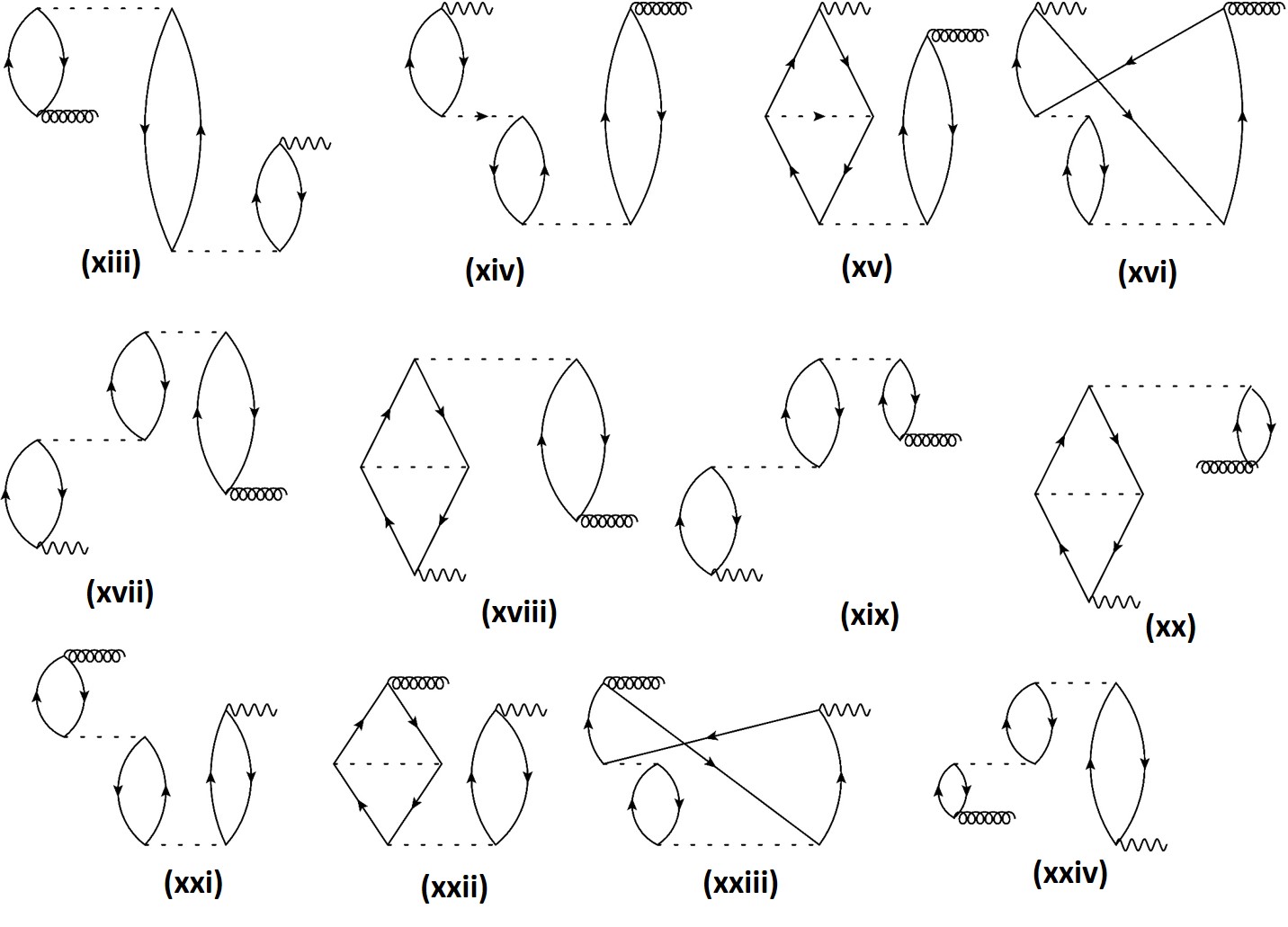}
\includegraphics[width=8.5cm,height=4.7cm]{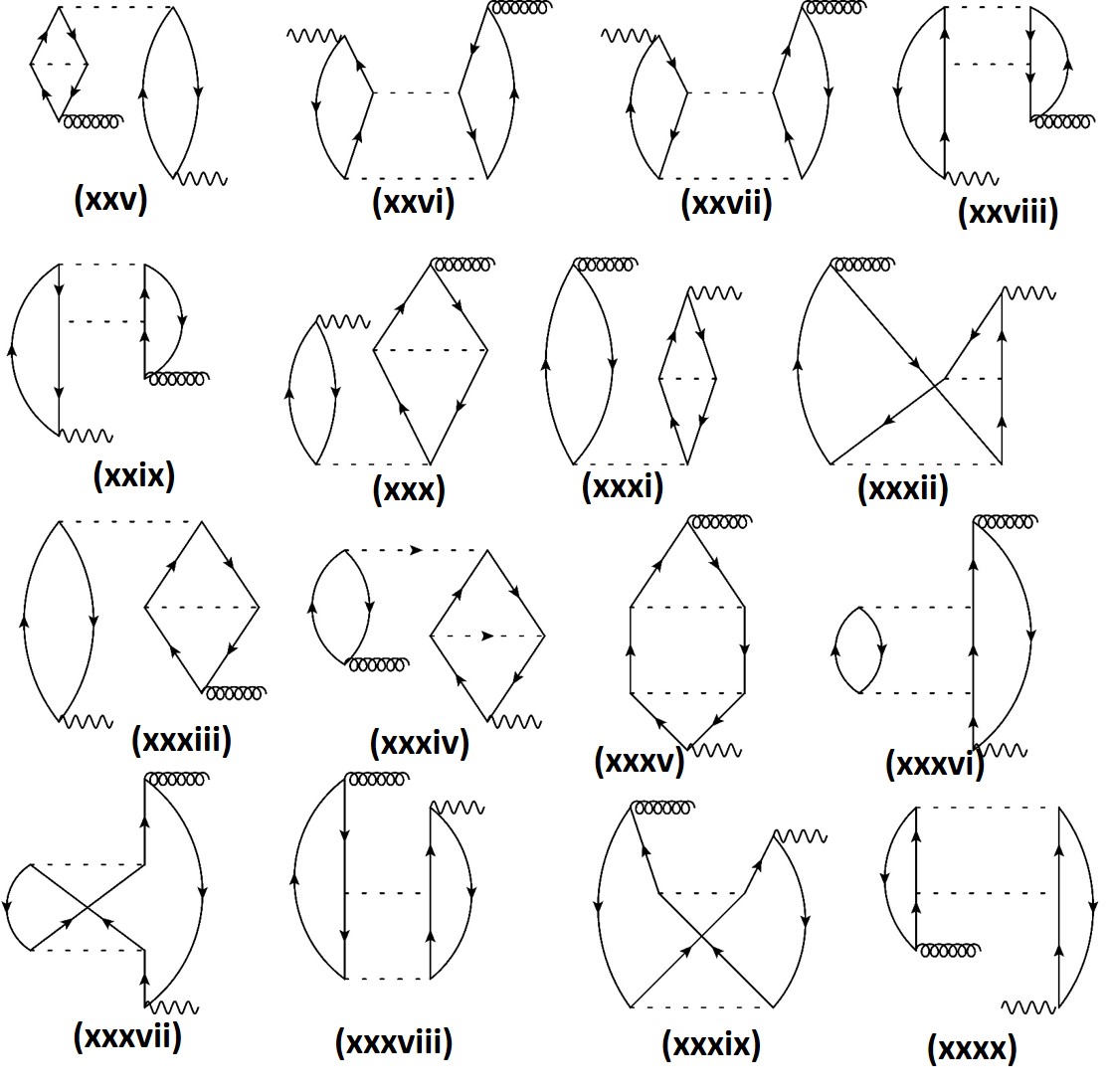}
\includegraphics[width=8.5cm,height=2.8cm]{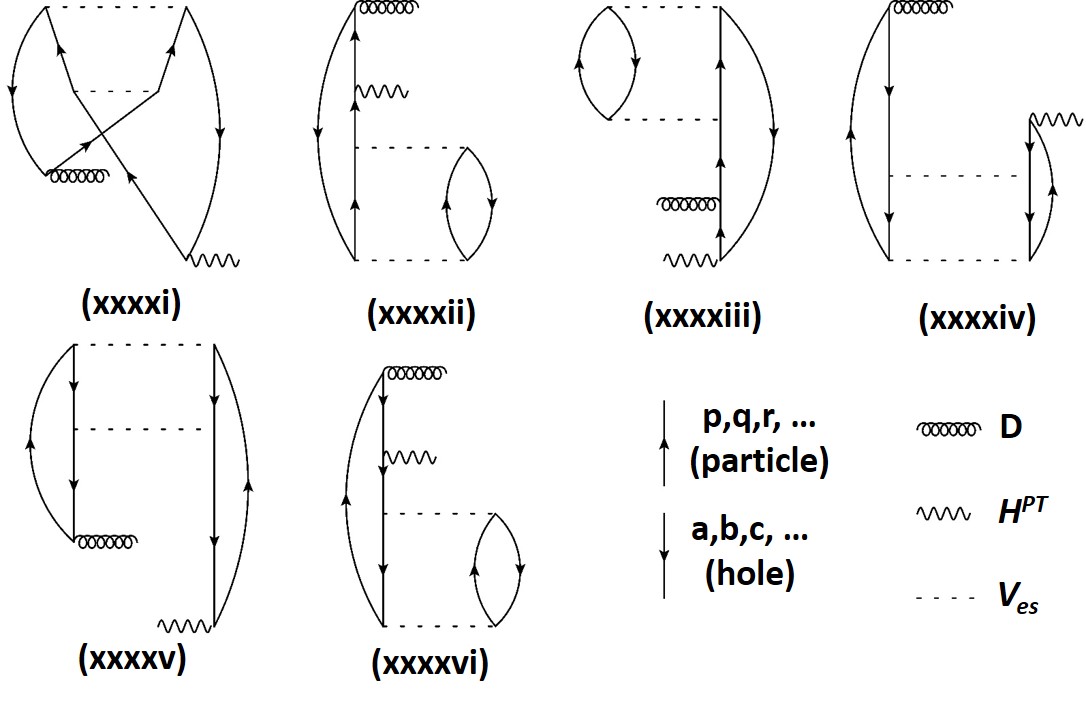}
\end{center}
\caption{Important Goldstone diagrams belonging to the MBPT(3) method. Diagram (i) and diagrams up to (vii) correspond to the DHF and 
MBPT(2) methods, respectively. Operators appearing from right to left in Eqs. (\ref{eq20}) and (\ref{eq21}) are shown 
from bottom to top. Lines with arrows up and down represent for the virtual and core orbitals, respectively. Symbols for different operators 
are shown at the end.}
\label{fig2}
\end{figure}

\begin{table}[t]
\caption{Explicit contributions to $\alpha_d$, ${\cal R_T}$, and ${\cal R_S}$ in units of $e a_0^3$, $\times 10^{-20} \langle 
\sigma\rangle |e|cm$ and $\times 10^{-17}[1/|e|fm^3]|e|cm$, respectively, from some of the important Goldstone diagrams 
of the MBPT(3) method. First one and along this the next six give DHF and MBPT(2) contributions, respectively.}
\begin{ruledtabular}
\begin{tabular}{lccc}
Diagrams & $\alpha_d $&  ${\cal R_T}$ &  ${\cal R_S}$ \\
\hline
 & & \\
Fig. \ref{fig2}(i) &  124.50  & $-0.71$  & $-0.42$  \\
Fig. \ref{fig2}(ii) &  $-46.95$  & $-0.02$   & $-0.01$     \\
Fig. \ref{fig2}(iii) &  92.82   & $-1.29$  & $-0.72$  \\
Fig. \ref{fig2}(iv) &  $-23.47$ & $-0.41$   & $-0.23$ \\
Fig. \ref{fig2}(v) & 8.91 & $-0.29$ & $-0.18$ \\
Fig. \ref{fig2}(vi) & $-23.47$  & 0.39  & 0.22 \\
Fig. \ref{fig2}(vii) & 8.91 & $-0.15$ & $-0.08$ \\   
Fig. \ref{fig2}(viii) & 25.03 & 0.16 & 0.09 \\
Fig. \ref{fig2}(ix) & $-35.26$ & 0.51 & 0.28 \\
Fig. \ref{fig2}(x) & $-35.26$ & 0.03 & 0.02 \\
Fig. \ref{fig2}(xi) & $-31.05$ & 0.42 & 0.23 \\
Fig. \ref{fig2}(xii) & 9.07 & $-0.11$ & $-0.06$ \\
Fig. \ref{fig2}(xiii) & 5.11 &  0.13 & 0.07 \\ 
Fig. \ref{fig2}(xiv) & 7.69 & 0.26 & 0.14 \\
Fig. \ref{fig2}(xv) & $-9.09$ & 0.04 & 0.03 \\
Fig. \ref{fig2}(xvi) & $-2.25$   &  0.10 & 0.06 \\
Fig. \ref{fig2}(xvii) & 6.49  & $-0.12$ & $-0.07$ \\
Fig. \ref{fig2}(xviii) & $-8.15$  &  0.13 & 0.07 \\
Fig. \ref{fig2}(xix) & 10.85 & 0.03 & 0.02 \\
Fig. \ref{fig2}(xx) & $-18.02$ & 0.33 & 0.18 \\
Fig. \ref{fig2}(xxi) & 6.49 & $-0.12$ & $-0.07$ \\
Fig. \ref{fig2}(xxii) & $-8.15$ & 0.12 & 0.07 \\
Fig. \ref{fig2}(xxiii) & $-2.22$ & 0.10 & 0.06 \\ 
Fig. \ref{fig2}(xxiv) & 7.69 &  $-0.15$ & $-0.08$ \\
Fig. \ref{fig2}(xxv) & $-9.09$ & 0.16 & 0.08 \\
Fig. \ref{fig2}(xxvi) & $-8.06$ & 0.14 & 0.07 \\
Fig. \ref{fig2}(xxvii) & $-8.00$ & 0.12 & 0.06 \\
Fig. \ref{fig2}(xxviii) & $-8.09$ & 0.14 & 0.08 \\
Fig. \ref{fig2}(xxix) & $-8.12$ & 0.13 & 0.07 \\
Fig. \ref{fig2}(xxx) & $-8.15$ & 0.12 & 0.07 \\
Fig. \ref{fig2}(xxxi) & $-9.09$ & 0.04 & 0.03 \\
Fig. \ref{fig2}(xxxii) & 3.59  &  $-0.11$ & $-0.07$ \\
Fig. \ref{fig2}(xxxiii) & $-9.09$ & 0.16 & 0.08 \\
Fig. \ref{fig2}(xxxiv) & $-8.18$ & 0.13 & 0.07 \\
Fig. \ref{fig2}(xxxv) & 70.93  & $-1.24$ & $-0.67$ \\
Fig. \ref{fig2}(xxxvi) & 15.72 & $-0.43$ & $-0.23$ \\
Fig. \ref{fig2}(xxxvii) & $-3.84$  & 0.10 & 0.06 \\
Fig. \ref{fig2}(xxxviii) & 11.83 & $-0.17$ & $-0.10$ \\
Fig. \ref{fig2}(xxxix) & 5.38 & $-0.11$ & $-0.07$ \\
Fig. \ref{fig2}(xxxx) & 11.83 & $-0.20$ & $-0.11$ \\
Fig. \ref{fig2}(xxxxi) & 5.38 &  $-0.09$ & $-0.05$ \\
Fig. \ref{fig2}(xxxxii) & $-11.67$ & $-0.25$ & $-0.14$ \\
Fig. \ref{fig2}(xxxxiii) & $-11.67$ & $-0.04$ & $-0.02$ \\
Fig. \ref{fig2}(xxxxiv) & 8.91  & $-0.11$ & $-0.06$ \\
Fig. \ref{fig2}(xxxxv) & 8.91 & $-0.14$ & $-0.07$ \\
Fig. \ref{fig2}(xxxxvi) & $-0.19$ & $-0.10$ & $-0.06$ \\
\end{tabular} 
\end{ruledtabular}
\label{tab2}
\end{table}

In Table \ref{tab1}, we present $\alpha_d$ and ${\cal R}$ values due to both the T-PT and NSM interactions in $^{171}$Yb by means of  
the earlier discussed many-body methods and compare them with the previously reported results \cite{dzuba7,dzuba9,dzuba2,latha,radziute}.
For convenience, we denote ${\cal R}$ values due to the T-PT and NSM interactions as ${\cal R_T}$ and ${\cal R_S}$, respectively. 
As can be seen, the DHF value of $\alpha_d$ and the CCSD result differ marginally giving an impression that the roles of the electron 
correlation effects in the evaluation of atomic wave functions in this atom are not very strong. However, analyzing results for this 
quantity from the MBPT(2), MBPT(3), and RPA methods indicate a different scenario. The MBPT(2) method gives a larger value, while the 
MBPT(3) method gives a lower value of $\alpha_d$ from the DHF method. The all order RPA method gives a very large value than all these 
methods and the all order CCSD method brings down this value drastically. It can be noted that the MBPT(2) method posses all the lowest 
order core-polarization effects and the MBPT(3) method accounts for the lowest order correlation effects that do not belong to the 
core-polarization effects, which are discussed elaborately below. Significant differences between the $\alpha_d$ values from the MBPT(2) and 
MBPT(3) methods suggest substantial contributions from these other than core-polarization effects and with the opposite sign than the 
core-polarization contributions. In the all order level, differences between the RPA and CCSD results imply the net contributions from the 
other than core-polarization effects. A preliminary experimental result on $\alpha_d$ of the ground state of Yb has been reported with large 
uncertainty as 142(36) a.u. \cite{miller}. However, many calculations on this quantities are carried out employing variants of many-body 
methods \cite{porsev,zhang,bksahoo,dzuba,safronova}. Most of these calculations are not consistent owing to large electron correlation effects 
associated with this atom. We had also obtained earlier this value using the CCSD method and was reported as 144.59(5.64) a.u. \cite{bksahoo}, 
considering only the linear terms of $\overline{D}^{(0)}$ in Eq. (\ref{eq38}). We find inclusion of contributions from the non-linear terms 
reduces this value because of which we give a slightly smaller value here. Nevertheless, the present result is in close agreement with most of 
the theoretical values and also with the central value of the reported experimental result. From this, we can hope that our CCSD method can 
also estimate the ${\cal R}$ values with reasonable accuracies.

\begin{table}[t]
\caption{Contributions to $\alpha_d$, ${\cal R_T}$, and ${\cal R_S}$ in units of $e a_0^3$, $\times 10^{-20} \langle \sigma\rangle |e|cm$ 
and $\times 10^{-17}[1/|e|fm^3]|e|cm$, respectively, from various CCSD terms (hermitian conjugate terms are included).}
\begin{ruledtabular}
\begin{tabular}{lccc}
CCSD term & $\alpha_d $&  ${\cal R_T}$ &  ${\cal R_S}$ \\
\hline
 & & \\
$DT_1^{(1)}$                &  160.55   & $-2.70$  & $-1.87$  \\
$T_1^{(0)\dagger}DT_1^{(1)}$&  $-11.58$  & $-0.04$    & $-0.01$     \\
$T_2^{(0)\dagger}DT_1^{(1)}$&  $-19.62$   & 0.68  &  0.41 \\
$T_1^{(0)\dagger}DT_2^{(1)}$&  1.06 &  $-0.03$  & 0.02 \\
$T_2^{(0)\dagger}DT_2^{(1)}$&  9.05 &  $-0.01$ &  $-0.10$    \\
& & \\
Higher                      &  $-3.96$    & 0.06  & 0.04 \\
\end{tabular} 
\end{ruledtabular}
\label{tab3}
\end{table}

 Though both the rank and parity of the E1 operator are same with the considered P,T-odd interaction operators, it can be clearly 
seen from Table \ref{tab1} that the trends of electron correlation effects in the evaluation of the $\alpha_d$ and ${\cal R}$ values are 
completely different in these properties. The CCSD results for ${\cal R}$ are almost three times larger than their corresponding DHF values. 
However, the values obtained in the MBPT(2) method are larger than the CCSD results and the MBPT(3) values are smaller than the MBPT(2) 
results. The RPA method in these cases also give very large values as compared to the CCSD results. One can notice that the electron correlation 
effects in the evaluation of ${\cal R_T}$ and ${\cal R_S}$ are also somewhat different. The CCSD value for ${\cal R_T}$ is smaller than the 
MBPT values, while it is larger in case of the NSM interaction. In Fig. \ref{fig1}, we plot the $(O-O_{DF})/O_{DF}$ contributions to $\alpha_d$ 
and ${\cal R}$ values with $O$ representing values from different many-body methods, which highlight the amount of electron effects that are 
being accounted for in the evaluation of these quantities through the respective methods. This clearly demonstrates that the role electron 
correlation effects play vital roles in determining the ${\cal R}$ values more than the $\alpha_d$ result. Few earlier calculations on 
${\cal R}$ are available using the configuration interaction (CI) \cite{angom}, RPA \cite{dzuba7,dzuba2,dzuba9,latha}, multi-configuration 
Dirac-Fock (MCDF) \cite{radziute} and a combined CI and MBPT methods \cite{dzuba9}. Our RPA values agree with the RPA results of Dzuba et al 
\cite{dzuba7,dzuba2}, but differ slightly from the RPA values reported by Latha and Amjith \cite{latha}. The CI, RPA and MCDF methods appear to 
overestimate the ${\cal R}$ values than the CCSD method. Similar trends of the correlation effects were also observed earlier in $^{199}$Hg
\cite{yashpal3,sahoo1} and in $^{225}$Ra \cite{yashpal4}. This clearly demands for employing a potential many-body method to evaluate the 
${\cal R}$ values with the reasonable accuracy so that they can be combined with the future experimental result of the $^{171}$Yb atom to infer 
more reliable limits on the $C_T$ and $S$ values.

\begin{figure}[t]
\begin{center}
\includegraphics[width=8.5cm,height=5.1cm]{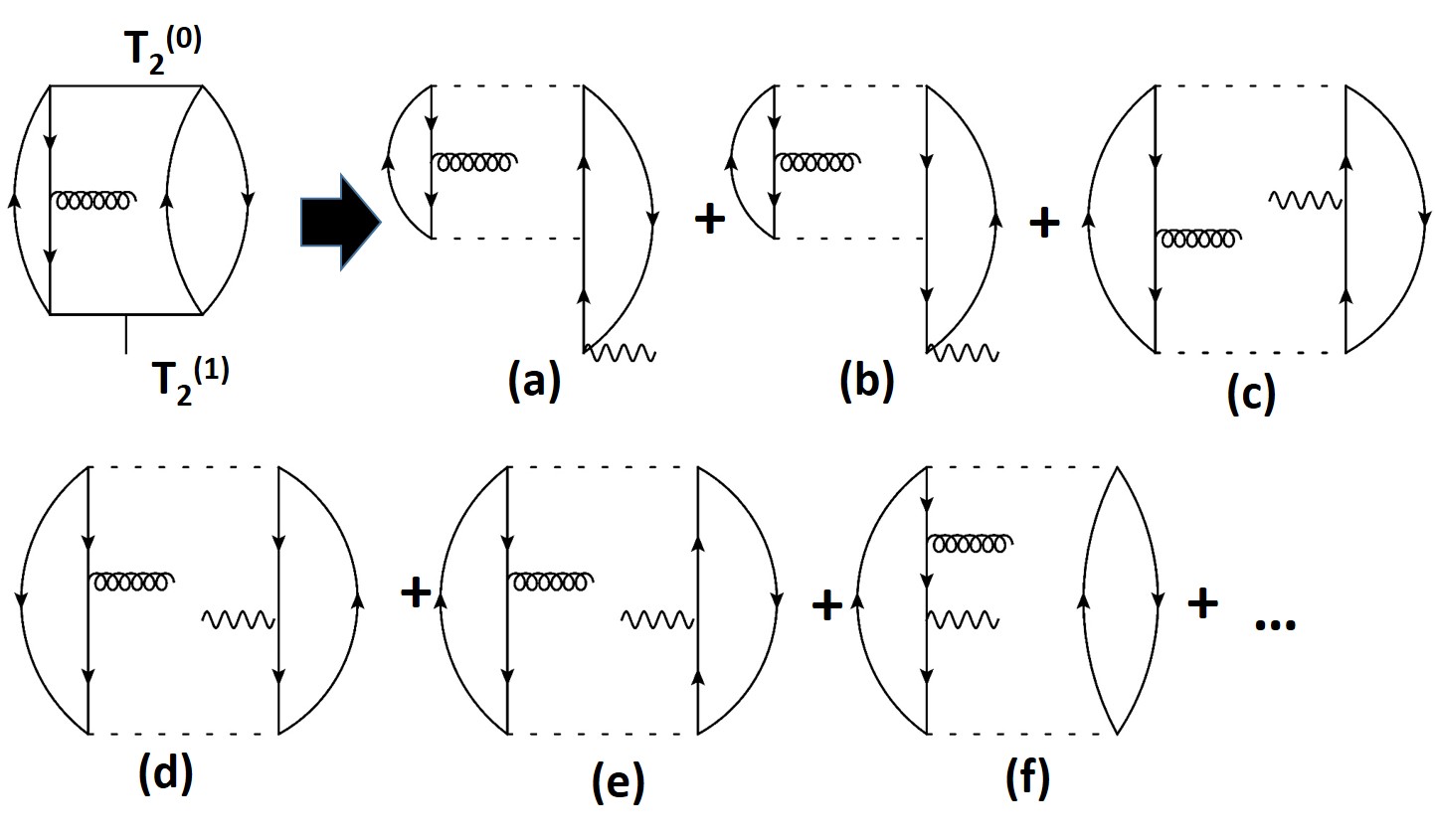}
\end{center}
\caption{Breakdown of one of the all order diagrams representing the $T_2^{(0)\dagger}DT_2^{(1)}$ term is shown explicitly for the 
comprehensive understanding of how other than the core-polarization effects are accounted for through the CCSD method in the evaluation of the 
$\alpha_d$ and ${\cal R}$ values.}
\label{fig3}
\end{figure}

\begin{figure}[t]
\begin{center}
\includegraphics[width=4.0cm,height=6.0cm]{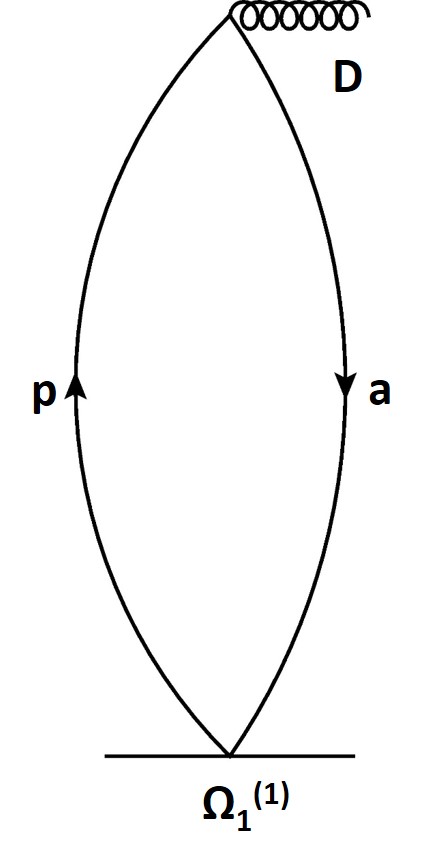}
\end{center}
\caption{The Goldstone diagram depicting singly excitation contributions to $\alpha_d$ and ${\cal R}$ by exciting a core orbital ``a'' 
to a virtual orbital ``p'' through the $\Omega^{(1)}$ operator. Here $\Omega^{(1)}$ represents for the first order perturbed 
operator in the DHF, MBPT(2) and RPA methods and the $T_1^{(1)}$ operator of the CCSD method.}
\label{fig4}
\end{figure}

After presenting the final results from various many-body methods, we now intend to analyze the roles of the electron correlation effects 
in the evaluation of the $\alpha_d$ and ${\cal R}$ values through various Goldstone diagrams of the MBPT, RPA and CCSD methods. In Fig.
\ref{fig2}, we show some of the important diagrams belonging to the MBPT(3) method. There are more than 200 diagrams appear in the MBPT(3) 
method, but we present here contributions only from the selective diagrams that contribute substantially. The first diagram of Fig. \ref{fig2}
represents for the DHF method and diagrams up to Fig. \ref{fig2}(vii) correspond to the MBPT(2) method. Individual contribution from these 
diagrams to $\alpha_d$ and ${\cal R}$ are given in Table \ref{tab2}. Some of the non-quoted diagrams also contribute in the similar orders
with slightly smaller values, but their contributions are not mentioned explicitly here to avoid a very long table. As can be seen from this 
table, magnitudes and signs of the contributions from various diagrams to $\alpha_d$ and ${\cal R}$ with respect to their respective final 
values exhibit different trends. Contributions from some diagrams to $\alpha_d$ are large, while they contribute tinier to ${\cal R}$. It 
is also found that some of the individual diagrams contribute as large as three-fourth of the total value to $\alpha_d$. Certain third order 
perturbative diagrams also contribute more than the second order diagrams to this property. Diagrams shown as Fig. \ref{fig2}(xi), (xii), 
(xxvi), (xxvii), (xxix), (xxxvi), (xxxxii), (xxxxiii), (xxxxiv), (xxxxv), (xxxxvi) etc. are some of the dominantly contributing MBPT(3) diagrams 
that represent for other than the core-polarization effects. These diagrams are solely responsible for bringing down the MBPT(3) values from the results 
obtained using the MBPT(2) method. They do not appear through the all order RPA method but appear in the CCSD method to all orders. This is 
the main reason why the CCSD results are found to be much smaller than the RPA values as quoted in Table \ref{tab1}. Comparing contributions 
to the ${\cal R}$ values from the T-PT and NSM interactions, we find they maintain a scaling among the contributions from each diagram. 
Contributions to the T-PT result are about two times larger than the NSM contributions for the individual diagram. In fact, some of the 
correlation contributions to $\alpha_d$ are found to have opposite sign than its DHF value, hence canceling out a large part of the 
correlation contributions to give the smaller net value. Contrary, the dominant contributions from the MBPT method to ${\cal R}$ have the same sign 
with their DHF values from the respective P,T-odd interactions. This is why enhancement in the ${\cal R}$ values from their DHF values are 
found to be much larger than the $\alpha_d$ result. It is also found that other than the core-polarization contributions are proportionally larger 
in the determination of the ${\cal R}$ values than the $\alpha_d$ result.

\begin{table*}[t]
\caption{Contribution from various atomic orbitals to $\alpha_d$, ${\cal R_T}$, and ${\cal R_S}$ in units of $e a_0^3$, $\times 10^{-20}
\langle \sigma\rangle |e|cm$ and $\times 10^{-17}[1/|e|fm^3]|e|cm$, respectively, through dominantly contributing singly excitations 
represented by Fig. \ref{fig4}. Values that are smaller in magnitudes are approximated to zero (mentioned as $\sim 0.0)$ and those marked 
in bold fonts highlight changes in the trends of the results from the DHF method after incorporating electron correlation effects through
different many-body methods. Contributions from the $6s$ and $p_{3/2}$ orbitals are quoted within two lines to demonstrate how they behave 
differently in the evaluation of the $\alpha_d$ and ${\cal R}$ values using the considered methods.}
\begin{ruledtabular}
\begin{tabular}{cc cccc cccc cccc}
Occupied & Virtual & \multicolumn{4}{c}{$\alpha_d$} &  \multicolumn{4}{c}{${\cal R_T}$} &  \multicolumn{4}{c}{${\cal R_S}$} \\
\cline{3-6} \cline{7-10} \cline{11-14} \\ 
 $a$            &     $p$  & DHF & MBPT(2) & RPA & $DT_1^{(1)}$  & DHF & MBPT(2) & RPA & $DT_1^{(1)}$  & DHF & MBPT(2) & RPA & $DT_1^{(1)}$  \\
\hline
 & & \\   
$4s$      & $11p_{1/2}$  &  $\sim 0.0$   & $\sim 0.0$    & $\sim 0.0$ & $\sim 0.0$ &  0.02     &  $-0.07$  & 0.03 &  0.03          & 0.01 & $-0.02$         & 0.01 & 0.01 \\
$3s$      & $12p_{1/2}$  &  $\sim 0.0$   & $\sim 0.0$    & $\sim 0.0$ & $\sim 0.0$ & 0.02      &  $-0.04$  & 0.02 &  0.02          & 0.01 & $-0.01$         & 0.01 & 0.01 \\
$4s$      & $12p_{1/2}$  &  $\sim 0.0$   & $\sim 0.0$    & $\sim 0.0$ & $\sim 0.0$ & 0.05      & $-0.14$   & 0.05 & 0.05           & 0.01 & $-0.03$         & 0.01 & 0.01\\
$5s$      & $10p_{1/2}$  &  $\sim 0.0$   & $\sim 0.0$    & $\sim 0.0$ & $\sim 0.0$ & 0.04      & 0.09        & 0.05 & 0.05           & 0.01 & 0.02              & 0.01 & 0.01 \\
$5s$      & $11p_{1/2}$  &  $\sim 0.0$   & $\sim 0.0$    & $\sim 0.0$  & $\sim 0.0$ & 0.07     &  0.01        & 0.08 & 0.08           & 0.02 & $\sim 0.0$    & 0.02 & 0.02 \\
$5s$      & $12p_{1/2}$  &  $\sim 0.0$   & $\sim 0.0$    & $\sim 0.0$ & $\sim 0.0$ & 0.02      & $-0.06$   & 0.02 & 0.02           &$\sim 0.0$  & $-0.01$& 0.01& 0.01 \\
$6s$      & $6p_{1/2}$    &  9.38             &  12.39          & 15.97         & 11.84         & $-0.22$ & $-0.29$  & $-0.78$& $-0.76$  & $-0.05$ & $-0.07$    & $-0.18$ & $-0.16$ \\
$6s$      & $7p_{1/2}$    &  22.54           &  28.94          & 37.35         & 29.90         &  $-0.69$ & $-0.88$ & $-{\bf 2.32}$ & $-{\bf 2.29}$ & $-0.17$ & $-0.21$    & $-{\bf 0.54}$ & $-{\bf 0.49}$ \\
$6s$      & $8p_{1/2}$    &  7.95             &  8.80            & 11.37         & 12.09         & $-0.52$  & $-0.57$  & $-{\bf 1.53}$ & $-{\bf 1.59}$& $-0.12$ & $-0.14$    & $-{\bf 0.37}$ & $-{\bf 0.35}$ \\
$6s$      & $9p_{1/2}$    &  0.20             &  0.20            & 0.04           & 0.36           & $-0.12$  &  0.05       & $-0.27$ & $-0.29$& $-0.03$ &  0.01       & $-0.07$ & $-0.07$ \\
$6s$      & $10p_{1/2}$  &  $\sim 0.0$   & $\sim 0.0$   & $\sim 0.0$  & $\sim 0.0$  & 0.02       &  0.11       & 0.03       & 0.03    & $\sim 0.0$ & 0.03    & 0.01 & 0.01 \\
\hline
$6s$      & $6p_{3/2}$    &  14.01           &  18.57          & 23.44         & 17.54         & $\sim 0.0$ &$\sim 0.0$ & 0.05  & 0.19    & $-0.05$ & $-0.07$   & $-0.16$ & $-0.13$ \\
$6s$      & $7p_{3/2}$    &  43.44           &  55.96          & {\bf 70.39}         & 57.35         & $\sim 0.0$ &$\sim 0.0$ &0.17   & {\bf 0.71}    & $-0.21$ & $-0.27$   & $-{\bf 0.63}$ & $-{\bf 0.53}$  \\
$6s$      & $8p_{3/2}$    &  18.73           &  20.99          & 25.86         & 28.03         &$\sim 0.0$ &$\sim 0.0$ & 0.06   & {\bf 0.48}     & $-0.21$ & $-0.21$   & $-{\bf 0.52}$ & $-{\bf 0.48}$ \\
\hline
$5p_{1/2}$ & $8s$         &  $\sim 0.0$   & $\sim 0.0$    &  $\sim 0.0$ &  $\sim 0.0$&  0.02       &$\sim 0.0$  & 0.02   &0.02 &$\sim 0.0$ &$\sim 0.0$  & 0.01 & $\sim 0.0$  \\
$5p_{1/2}$ & $9s$         &  0.03             &  $\sim 0.0$   &  $-0.03$     & $\sim 0.0$ &  0.09        & $-0.02$     & 0.13    &0.11    & 0.02    & $\sim 0.0$ & 0.03 & 0.03 \\
$4p_{1/2}$ & $10s$       &  $\sim 0.0$    & $\sim 0.0$   & $\sim 0.0$  &  $\sim 0.0$ &  0.01       & $-0.02$     & 0.01    &0.01    &$\sim 0.0$  & $-0.01$ &$\sim 0.0$  & $\sim 0.0$  \\
$5p_{1/2}$ & $10s$       &  0.03              &  $-0.01$      &   $-0.04$    & $\sim 0.0$   &  0.19       & $-0.09$     & {\bf 0.26}     & {\bf 0.24}   &0.05    & $-0.02$        & {\bf 0.07} & {\bf 0.06} \\
$4p_{1/2}$ & $11s$       &  $\sim 0.0$     & $\sim 0.0$    &  $\sim 0.0$ & $\sim 0.0$ &  0.04       &  $-0.12$      & 0.04 & 0.04              & 0.01 & $-0.03$  & 0.01 & 0.01 \\
$5p_{1/2}$ & $11s$       &  $\sim 0.0$     &  $\sim 0.0$   &  $\sim 0.0$ & $\sim 0.0$ &  0.09       & $-0.12$     & 0.10 & 0.10              & 0.02  & $-0.03$ & 0.03 &0.02 \\
$3p_{1/2}$ & $12s$       &  $\sim 0.0$     &   $\sim 0.0$  &  $\sim 0.0$ & $\sim 0.0$ &  0.01       &  $-0.02$    & 0.02    & 0.02   &$\sim 0.0$  & $-0.01$ & $\sim 0.0$  & $\sim 0.0$  \\
$4p_{1/2}$ & $12s$       &  $\sim 0.0$     &   $\sim 0.0$  &  $\sim 0.0$ & $\sim 0.0$ &  0.03        &  $-0.08$   & 0.03 & 0.03            &0.01     & $-0.02$ & 0.01 & 0.01 \\
$5p_{3/2}$ & $8s$         &   0.03             &  $\sim 0.0$    &  $-0.03$     &  0.01       &$\sim 0.0$   & $\sim 0.0$&$\sim 0.0$ &$\sim 0.0$  & 0.01    & $\sim 0.0$  & 0.01 & 0.01 \\
$5p_{3/2}$ & $9s$         &   0.11             &  $-0.03$        &  $-0.13$     &   0.02       &$\sim 0.0$   & $\sim 0.0$ &  0.02   &  0.01              &  0.04   & $-0.01$      & 0.07 &0.06 \\
$4p_{3/2}$ & $10s$       &   $\sim 0.0$    &  $\sim 0.0$   &  $\sim 0.0$ &  $\sim 0.0$&$\sim 0.0$ & $\sim 0.0$ &$\sim 0.0$ &$\sim 0.0$ & 0.01 & $-0.01$         & 0.01 & 0.01 \\
$5p_{3/2}$ & $10s$       &   0.10             & $-0.06$         &  $-0.13$     &   $\sim 0.0$&$\sim 0.0$& $\sim 0.0$  &  0.03    &0.02               & 0.08 & $-0.05$        & {\bf 0.12} & {\bf 0.11} \\
$4p_{3/2}$ & $11s$       &   $\sim 0.0$     &  $\sim 0.0$  &  $\sim 0.0$ &  $\sim 0.0$  & $\sim 0.0$ &$\sim 0.0$&$\sim 0.0$ & $\sim 0.0$ & 0.02 & $-0.05$       & 0.01 & 0.02 \\
$5p_{3/2}$ & $11s$       &   $\sim 0.0$     &  $-0.01$       &  $\sim 0.0$ &  $\sim 0.0$  & $\sim 0.0$ &$\sim 0.0$&$\sim 0.0$  &$\sim 0.0$ & 0.03 & $-0.05$      & 0.03 & 0.04\\
$3p_{3/2}$ & $12s$       &   $\sim 0.0$     &  $\sim 0.0$  & $\sim 0.0$   &  $\sim 0.0$ & $\sim 0.0$  &$\sim 0.0$&$\sim 0.0$ &$\sim 0.0$ &  0.01 & $-0.01$     &$\sim 0.0$  & 0.01 \\
$4p_{3/2}$ & $12s$       &   $\sim 0.0$    &  $\sim 0.0$   & $\sim 0.0$   &  $\sim 0.0$ &  $\sim 0.0$ &$\sim 0.0$&$\sim 0.0$ &$\sim 0.0$ &  0.01 & $-0.03$     & 0.01 & 0.01 \\
$5p_{3/2}$ & $12s$       &   $\sim 0.0$    &   $\sim 0.0$  & $\sim 0.0$   &  $\sim 0.0$ & $\sim 0.0$  &$\sim 0.0$&$\sim 0.0$ &$\sim 0.0$ & $\sim 0.0$  & $-0.01$ & $\sim 0.0$  & $\sim 0.0$  \\
\end{tabular} 
\end{ruledtabular}
\label{tab4}
\end{table*}

 In Table \ref{tab3}, we present contributions from the individual CCSD terms. This shows the dominant contributions come from the 
$DT_1^{(1)}$ term followed by $T_2^{(0)\dagger}DT_1^{(1)}$. Contribution from $T_1^{(0)\dagger}DT_1^{(1)}$ to $\alpha_d$ is also quite 
large, however its contributions to ${\cal R}$ are very small. It can be noticed from Fig. \ref{fig1} that contributions from most of these
diagrams from the MBPT(3) method are accounted for in the $DT_1^{(1)}$ term through the RCC formulation and the rest arise through the 
$T_2^{(0)\dagger}DT_1^{(1)}$ term. This is the reason why $DT_1^{(1)}$ and $T_2^{(0)\dagger}DT_1^{(1)}$ terms have major shares to the CCSD 
results as demonstrated in Table \ref{tab3}. In addition to the above, contributions coming through $T_1^{(0)\dagger}DT_1^{(1)}$ are also 
from the singly excitations in the configuration space and offer significant shares to the final results. It can be noticed from the above 
table that a substantial amount of contributions also come through the $T_2^{(0)\dagger}DT_2^{(1)}$ term. Breakdown of one of its diagrams 
into some of the MBPT(3) diagrams are shown in Fig. \ref{fig3}. As seen from this figure, all the contributions arising through the 
$T_2^{(0)\dagger}DT_2^{(1)}$ term are due to other than the core-polarization effects and they are important in determining 
the $\alpha_d$ value, while they contribute relatively smaller in the evaluation of the ${\cal R}$ values. These MBPT(3) diagrams were not 
shown explicitly in Fig. \ref{fig1} as each of these diagrams contribute small, but they add up to a sizable amount in the CCSD method. In 
the above table, we also quote contributions from the remaining terms of the CCSD method as ``Higher'', because they correspond to higher
order correlation effects and arise through the non-linear terms such as the $T_1^{(0)\dagger}DT_1^{(0)}T_1^{(1)}$, 
$T_2^{(0)\dagger}DT_2^{(0)}T_1^{(1)}$, etc. terms. Most of these contributions are due to other than the core-polarization effects.

  The Goldstone diagram depicting the $D \Omega^{(1)}$ term with the approximation for $\Omega^{(1)}$ as the first order perturbed 
operator in the DHF, MBPT(2) and RPA methods and the $T_1^{(1)}$ operator of the CCSD method is shown in Fig. \ref{fig4}. This 
represents the dominantly contributing singly excited configurations, which we have represented replacing a core orbital (a), denoted 
with a line pointing down arrow, by a virtual orbital (p), denoted with a line with upward arrow, at various levels of approximations. 
Contributions from this diagram at the DHF, MBPT(2), RPA and CCSD level are listed in Table \ref{tab4} only from the large contributing 
orbitals. As can be seen from this table, contributions from various orbitals to $\alpha_d$ and ${\cal R}$ values are different. In the 
determination of $\alpha_d$, only the $6s$ and $p_{1/2,3/2}$ orbitals play all the roles. This trend also shows why and how the RPA result 
for $\alpha_d$ becomes very large, particularly through the $6s-7p_{3/2}$ orbitals. It exhibits that the core-polarization effects are 
changing contributions from these orbitals very strongly at the MBPT(2) and RPA level of approximations, while other types of correlation  
effects coming through the $DT_1^{(1)}$ RCC term revamp these orbitals further to bring these values down. It can also be seen that the 
$6s-p_{3/2}$ orbitals contribute more than the $6s-p_{1/2}$ orbitals to this quantity. Comparison with the $\alpha_d$ results, the 
correlation effects affect more strongly to the atomic orbitals in the evaluation of the ${\cal R}$ values. Again, the $6s-p_{1/2}$ orbitals 
contribute predominantly in the evaluation of ${\cal R}$ than the $6s-p_{3/2}$ orbitals. In fact, contributions from the $6s-p_{3/2}$ to 
these quantities at the DHF values are negligibly small and other than the core-polarization effects through the $DT_1^{(1)}$ RCC term modified 
these orbitals drastically to give a quite significant contributions to the final results. To highlight the same, results only from the
$6s-p_{3/2}$ orbitals are put in between two lines of the above table. It can also be noticed that the $6s$ and $p_{1/2,3/2}$ orbitals 
contribute differently to the ${\cal R_T}$ and ${\cal R_S}$ values at various levels of approximations in the many-body methods. In contrast 
to the $\alpha_d$ value, some of the high-lying orbitals also contribute substantially to these quantities as these continuum have large 
overlap over the nuclear region. We have marked in bold fonts to some of the quoted contributions from few specific orbitals to bring into notice 
how the electron correlation effects modifies these orbitals unusually large in $^{171}$Yb for studying atomic properties. This implies that 
it is important to consider a potential many-body method to determine the ${\cal R}$ values in this atom. It also suggests testing accuracies
of the $\alpha_d$ value cannot justify accuracies of the ${\cal R}$ values absolutely, but it can only assure validity of the calculations to 
some extent.

\section{Conclusion}
 
Roles of electron correlation effects in the determination of dipole polarizability and electric dipole moment due to parity and time reversal 
symmetry violations considering the tensor-pseudotensor interactions between the electrons with the nucleus and electrons with the nuclear 
Schiff moment in the $^{171}$Yb atom are analyzed. For this purpose, relativistic many-body methods over the Dirac-Hartree-Fock wave 
function at the approximations of the second and third order many-body perturbation theories, random phase approximation and coupled-cluster 
method with singles and doubles excitations are employed. Contributions from the core-polarization effects and other possible correlation  
interactions are investigated categorically from the differences of the random phase approximation and coupled-cluster calculations. To fathom 
the origin of these differences, contributions in terms of the important Goldstone diagrams appearing through the second-order and third-order 
perturbative methods are given. Moreover, contributions from different orbitals at various levels of approximations in the many-body methods 
listed for the comprehensive understanding of propagation of the electron correlation effects in the above atom through these orbitals to the 
considered properties that have distinct radial behaviors. This suggests that accuracies in the calculated electric dipole moments in atoms 
cannot be really determined from the dipole polarizability calculations. On the grounds of physical effects that are being embodied in the 
calculations, the values obtained employing our coupled-cluster method are more accurate and they can be used to infer reliable limits 
on the tensor-pseudotensor coupling constant between the electrons and nucleus and nuclear Schiff moment of the $^{171}$Yb atom when its 
experiment comes to fruition.

\section*{Acknowledgement}

This work was partly supported by the TDP project of Physical Research Laboratory (PRL), Ahmedabad and the computations were carried out 
using the Vikram-100 HPC cluster of PRL.

\end{document}